\documentclass[11pt]{article}
\usepackage{amssymb}

\newcommand{\mod}{\ \ {\rm mod}\ }
\newcommand{\bdm}{\begin{displaymath}} \newcommand{\edm}{\end{displaymath}}
\newcommand{\beq}{\begin{equation}}    \newcommand{\enq}{\end{equation}}
\newcommand{\bea}{\begin{eqnarray}}    \newcommand{\eea}{\end{eqnarray}}
\newcommand{\barr}{\begin{array}}      \newcommand{\earr}{\end{array}}

\newcommand{\mco}{\multicolumn}      \newcommand{\tx}{\textstyle}
\newcommand{\tfr}{\textstyle\frac}
\newcommand{\ny}{\nonumber}          \newcommand{\tf}{\tx\frac} 
\newcommand{\Lb}{\left[}             \newcommand{\Rb}{\right]}       
\newcommand{\lb}{\left\{}            \newcommand{\rb}{\right\}}      
\newcommand{\lk}{\left(}             \newcommand{\rk}{\right)}       

\newcommand{\cA}{{\cal A}}           
\newcommand{\cB}{{\cal B}}           \newcommand{\caC}{{\cal C}}
                 
\newcommand{\cF}{{\cal F}}                      
\newcommand{\cH}{{\cal H}}           
           \newcommand{\cN}{{\cal N}}
\newcommand{\cP}{{\cal P}}                          
\newcommand{\cZ}{{\cal Z}}           \newcommand{\Csf}{\mathsf{C}} 
\newcommand{\Zmb}{\mathbb{Z}}         
\newcommand{\TQ}{\widetilde{Q}}      

\newcommand{\alp}{\alpha}             
           \newcommand{\om}{\omega}
            
\newcommand{\SiS}{\mbox{sin}}        
\newcommand{\hs}{\hspace*{1cm}}      \newcommand{\hq}{\hspace*{6mm}}  
\newcommand{\hx}{\hspace*{3mm}} 
\newcommand{\PQL}{P_Q^{(L)}}         \newcommand{\Pic}{\Pi_Q^{(L)}}
\newcommand{\Pdd}{\Pi^{(3)}}

\newcommand{\hal}{{\tf{1}{2}}}       \newcommand{\pid}{\frac{\pi}{3}}

\newcommand{\RAa}{\:\Rb_{-\frac{2}{3},-\frac{1}{3}}} 
 
\newcommand{\RBa}{\:\Rb_{-\frac{1}{3}, \frac{1}{3}}}
\newcommand{\RBb}{\:\Rb_{ \frac{1}{3},-\frac{1}{3}}}
\newcommand{\RCa}{\:\Rb_{-\frac{2}{3}, \frac{2}{3}}}
\newcommand{\RCb}{\:\Rb_{ \frac{2}{3},-\frac{2}{3}}}
\newcommand{\RDa}{\:\Rb_{ \frac{1}{3}, \frac{2}{3}}}
\newcommand{\REa}{\:\Rb_{ \frac{2}{5},-\frac{2}{5}}}
\newcommand{\RGa}{\:\Rb_{ \frac{3}{5},-\frac{3}{5}}}
\newcommand{\RFa}{\:\Rb_{ \frac{1}{5},-\frac{1}{5}}}

\newcommand{\rnc}{\renewcommand}     

\begin{document}

\begin{center}
{\LARGE\bf The superintegrable chiral Potts quantum chain and generalized
 Chebyshev polynomials}
\\[1cm]
{\Large\bf G. von Gehlen$^a$ and Shi-shyr Roan$^b$}\\[5mm]
$^a$Physikalisches Institut der Universit\"at Bonn\\
  Nussallee 12, 53115 Bonn, Germany\\{\it e-mail: gehlen@th.physik.uni-bonn.de}\\[5mm]
 $^b$Institute of Mathematics, Academia Sinica\\ 
 Taipei, Taiwan\\{\it e-mail: maroan@ccvax.sinica.edu.tw}
\end{center}
\vspace*{5mm}

\begin{abstract}
Finite-dimensional representations of Onsager's algebra are characterized by
the zeros of truncation polynomials. The $Z_N$-chiral Potts quantum chain 
hamiltonians (of which the Ising chain hamiltonian is the $N=2$ case) are the
main known interesting representations of Onsager's algebra and the 
corresponding polynomials have been found by Baxter and Albertini, McCoy 
and Perk in 1987-89 considering the Yang-Baxter-integrable 2-dimensional 
chiral Potts model. 
We study the mathematical nature of these polynomials. We find that for 
$N\ge 3$ and fixed charge $Q$ these don't form classical orthogonal sets 
because their pure recursion relations have at least $N+1$-terms. However, 
several basic properties are very similar to those required for orthogonal 
polynomials. The $N+1$-term recursions are of the simplest type: like for 
the Chebyshev polynomials the coefficients are independent of the degree. 
We find a 
remarkable partial orthogonality, for $N=3,\:5$ with respect to Jacobi-, and 
for $N=4,\:6$ with respect to Chebyshev weight functions. The separation 
properties of the zeros known from orthogonal polynomials are violated only 
by the extreme zero at one end of the interval.      
\end{abstract}


\section{Introduction}
Onsager's algebra (OA) has played a crucial role in the early history of 
exactly solvable statistical systems: it served as the tool for the first 
solution of the two-dimensional Ising model \cite{Onsa}. However, since 
soon after many other methods for solving the Ising model have been invented, 
for many years Onsager's algebra received little attention. It  
reappeared when the superintegrable $\Zmb_N$-symmetrical chiral Potts quantum
chain (SCPC) was introduced as a natural generalization of the transverse 
Ising quantum chain \cite{HKDN,GeRi}, and was shown to be integrable 
\cite{GeRi} because it forms representations of the OA (or the closely 
related Dolan-Grady algebra \cite{DoGr,Pe87}).
The actual solution for the lowest energy eigenvalues of the SCPC 
by Baxter \cite{Ba88} and Albertini, McCoy and Perk \cite{Al89} did not 
use the representation theory of Onsager's algebra but rather inversion 
relations, or, more generally, functional relations for the transfer 
matrix of the 2-dimensional Yang-Baxter integrable chiral Potts 
model \cite{AuYa,BaAu}.
In 1990 B.Davies \cite{Da90} constructed finite-dimensional 
representations of Onsager's algebra. In his approach zeros 
of truncation polynomials play a crucial role and the representations
are determined by a set of zeros and by $sl(2,C)$ representations.
While still it is not known how to obtain the Onsager algebra truncation 
polynomials directly from the SCPC hamiltonian, equivalent 
polynomials have been obtained from the 2-dimensional superintegrable 
Potts model functional relations in \cite{Ba88,Al89}.

Given the central role of the truncation polynomials for models solvable
due to Onsager's algebra, it seems desirable to improve our understanding of
their mathematical properties. In the Ising case, after a simple mapping,
these are Chebyshev polynomials (a fact implicit e.g. in \cite{Ba82}). 
For $N \ge 3$ we find many properties reminiscent of, 
but not in full agreement with those of orthogonal polynomials. 
Looking into recursion relations, differential equations, zero distributions 
and eventual weight functions for orthogonality, 
we find simple structures and in particular, a remarkable partial
orthogonality with Jacobi-weight functions.

In the next Section we review Onsager's algebra, the appearance of the
polynomials and Baxter's results for the $\Zmb_N$-polynomials. In Sec.3 we
discuss three different versions of the polynomials with different locations 
of the zeros and check how these look in the Ising case. Sec.4 gives some
properties of the $\Zmb_3$-polynomials with the zeros on the negative
real axis. Sec.5 gives our main results, which are for the polynomials
with zeros in the interval $-1<c<+1$: recursion relations, differential 
equations, zeros separation and, finally, partial orthogonality. 
Sec.6 contains our conclusions.

\section{Finite dimensional representations of Onsager's algebra} 
\subsection{Onsager's algebra}
Onsager's algebra $\cA$ is formed \cite{Onsa} from elements 
$A_m,\,G_l,\;m\in \Zmb,\;l\in {\mathbb N},\;\:l\ge m,$ 
satisfying 
\beq   [A_l,A_m] = 4\,G_{l-m};\hq
[G_l,A_m]= 2\,A_{m+l}-2\,A_{m-l};\hq  [G_l,G_m]= 0. 
\label{Ons}\end{equation}
Eqs.(\ref{Ons}) imply an infinite set of constraints:
\beq                [A_{m+1},A_m]=[A_m,A_{m-1}]      \end{equation}
and the existence of the infinite set of commuting operators $Q_m$:
\beq  Q_m=\hal\lk A_m+A_{-m}+k(A_{m+1}+A_{-m+1})\rk;\hs 
     \Lb\: Q_l,\:Q_m\Rb=0.           \label{consc}     \end{equation}   
where $k$ is a parameter which usually is taken to be real.
$A_0$ and $A_1$ generate $\cA$ if they satisfy the Dolan-Grady-relations
\beq  
\Lb A_0,\Lb A_0,\Lb A_0,A_1\Rb\,\Rb\,\Rb\,=\,16\Lb A_0,A_1\Rb;
\hs\;\Lb A_1,\Lb A_1,\Lb A_1,A_0\Rb\,\Rb\,\Rb\,=\,16\Lb A_1,A_0\Rb.
\label{DG} \end{equation}
Finite dimensional representations of $\cA$ are obtained 
\cite{Da90,Ro91,DaRo} if 
the $A_m$ satisfy a pure finite recurrence or difference equation: 
\beq      \sum_{k=-n}^n\;\alpha_k\,A_{k-l}\;=0        \end{equation}
(implying the same equation for the $G_l$).
For solving this relation B.Davies \cite{Da90} introduced the polynomial
\beq   \cF(z)=\sum_{k=-n}^n\alpha_k\:z^{k+n}.     \label{zpoly}\end{equation} 
From $\cA$ it follows that  
$\alpha_k$ is either even or odd in $k$: $\;\alpha_k=\pm\alpha_{-k}$, so that
\beq \cF(z)=\pm z^{2n+1}\cF(1/z) \label{srcf}\end{equation} and
the zeros of $\cF(z)$ come in reciprocal pairs $z_j,\;z_j^{-1}\;\:
(j=1,\ldots,n$). The $A_m$ and $G_m$ can be now be expressed in terms of a 
set of operators $\:E_j^\pm,\;H_j$: 
\beq 
A_m=2\sum_{j=1}^n\:\lk z_j^m\,E_j^+ +z_j^{-m}
 E_j^-\rk;\hs G_m\:=\:\sum_{j=1}^n\:\lk z_j^m-z_j^{-m} \rk \:H_j 
\end{equation}
which from $\cA$ obey $sl(2,C)$-commutation rules:
\beq [E_j^+,E_k^-]\:=\:\delta_{jk}\:H_k; \hs\hx
    [H_j,\:E_k^\pm]\:=\:\pm 2\,\delta_{jk}\:E_k^\pm. 
\end{equation}
So $\cA$ is isomorphic to a subalgebra of the loop algebra of
a direct sum of $sl(2,C)$ algebras.

We shall be interested in the eigenvalues of hermitian 
hamiltonians $\cH\sim A_0+k A_1$ ($\cH\sim Q_0$ of eq.(\ref{consc})), 
with parameter $k$ and $A_0$ and $A_1$ satisfying (\ref{DG}).
Write $E_j^\pm=J_{x,j}\,\pm\,iJ_{y,j}$,
then in a representation $\cZ(n,s)$ characterized by $z_1,\ldots,z_n$ and by 
a spin-$s$ representation $\vec{J}_j^{(s)}$ of all the $\vec{J}_j$, we have
\bea 
(A_0+k A_1)_{\cZ(n,s)}&=&2\,\sum_{j=1}^n\lb (2+k\,(z_j+z_j^{-1}))J_{x,j}^{(s)}
            +i(z_j-z_j^{-1})\,J_{y,j}^{(s)}\rb \ny\\
  &=&4\,\sum_{j=1}^n\,\sqrt{1+2k\,c_j+k^2}\;{J'}_{x,j}^{(s)}\ny\eea
where ${J'}_{x,j}$ is a $SU(2)$-operator and
\beq c_j\:=\;\cos{\theta_j}\;=\:\hal(z_j+z_j^{-1})   \end{equation}
with $\theta_j$ real for hermitian $\cH$.

\subsection{The superintegrable chiral Potts quantum chain}
The main example of hamiltonians $\cH$ satisfying the above conditions
(for generalisations see \cite{AhnS,IvUg}) are the 
$\Zmb_N$-SCPC-hamiltonians \cite{HKDN,GeRi} given by
\beq
\cH^{(s)}=-\,\sum_{j=1}^L\sum_{l=1}^{N-1}\frac{2}{1-\omega^{-l}}\lk X_j^l
      \;+\: k\:Z_j^l Z_{j+1}^{N-l}\rk      \label{SCP}  \end{equation} 
where $Z_j$ and $X_j$ are $\Zmb_N$-spin operators acting in the vector spaces 
${\mathbb{C}}^N$ at the sites $j\;$ $(j=1,\ldots,L)$: 
\beq Z_i\: X_j\:=\:X_j\:Z_i\:\omega^{\delta_{i,j}};\hs\hs 
  Z_j^N= X_j^N=1; \hs\hs \omega=e^{2\pi i/N}.   \end{equation}    
The normalization agrees with (\ref{DG}) if we write
$\;\cH^{(s)}=-\hal\,N(A_0+k A_1).$ 

In order to find the representations $\cZ(n,s)$ corresponding to the various 
sectors of (\ref{SCP}), no direct way is known. 
       Albertini {\it et al.} \cite{Al88} have recursively found the low-$k$ 
       finite-size ground state energy using only eq.(\ref{EE}) below and 
	   the order $k$-approximation of (\ref{SCP}), but this
       has not lead to a closed formula for the polynomials. 
However, these polynomials have been derived by solving the
two-dimensional chiral Potts model \cite{AuYa,BaAu}, which for the
"superintegrable" choice of parameters\footnote{"Superintegrable":
if integrable {\it both} due to Onsager's algebra and a 
Yang-Baxter equation.} contains the 
hamiltonian (\ref{SCP}) as a transfer matrix derivative. 
        First, using an inversion relation, Baxter \cite{Ba88} 
		found one low-lying
sector which for $k\rightarrow 0$ contains the ground state. 
Then, for the $\Zmb_3$-case, using functional relations for the 
transfer matrix, Albertini, McCoy and Perk \cite{Al89} obtained also the 
excited level sectors which involve solutions of Bethe-equations. 
Finally, Baxter \cite{Ba94}, exploiting functional relations derived
from the relation of the 2-dimensional integrable chiral Potts model to the 
six-vertex model \cite{BaSt}, 
solved the general $\Zmb_N$-superintegrable case. \footnote{For reviews
see e.g. \cite{MC90},\cite{AuPe}}. For all sectors of (\ref{SCP}) the spin 
representation turns out to be $s=\hal$ or the trivial one.
Accordingly, all eigenvalues of the hamiltonians (\ref{SCP}) have the form 
\beq E^{(s)}\,=-\frac{N}{2}\lb\,a\,+\,b\,k\,+\,4\,\sum_{j=1}^n m_j\,
 \sqrt{1+2\,k\cos{\theta_j}\,+\,k^2}\,\rb;\hs m_j=\pm\hal.  
\label{EE} \end{equation} 
$a$ and $b$ are integers originating from the trivial representation.

In this talk, we will consider only the low-lying 
(no Bethe-excitations) sector polynomials of Baxter \cite{Ba88,Ba94}.
Baxter's polynomials are written in the variable $t$ or $s=t^N$ which
is related to $z$ or $c=\cos{\theta}$ by 
\beq \frac{z+z^{-1}}{2}=c=\frac{1+s}{1-s} \label{zcs}\end{equation} 
The polynomials are (for details we refer to Baxter's paper
 \cite{Ba94}, for simplicity we put there $P_b=0$):
\beq \PQL(s)=\frac{1}{N}\:\sum_{j=0}^{N-1}
         \lk\frac{1-t^N}{1-\om^j\,t}\rk^L(\om^j\,t)^{-P_a};
\hs -P_a=Q+r+L \mod N. \label{Bscc}\end{equation}   
Here $Q$ denotes the $\Zmb_N$-charge sector ($Q=0,1,\ldots,N-1$), 
and $r$ ($r=0,1,\ldots,N-1$) labels the boundary 
condition to (\ref{SCP}) defined by $Z_{L+1}=\om^r Z_1.$ Since in (\ref{Bscc})
$Q$ and $r$ appear only in the sum $Q+r$, we can restrict us to consider the 
periodic case $r=0$.
The degree of the polynomials $\PQL(s)$ in the variable $s$ is 
\beq b_{L,Q}=\left[\frac{(N-1)L-Q}{N}\right] \label{bdi}\end{equation}
where $[x]$ denotes the integer part of $x$. 
Inversion of the variable $s$ leads to simple relations: 
\beq 
\PQL(s)\:=\:s^{b_{L,Q}}\:P_{N-(Q+L)}^{(L)}(s^{-1}), \label{srcp}\end{equation}
where the charge index $\:N-(Q+L)\:$ is understood $\!\!\mod N$. 
These relations correspond to the self-reverse property (\ref{srcf}) of the 
interpolation polynomials.

Considering sequences of these polynomials for fixed $Q$ and 
$L\in \mathbb{N}$,
from (\ref{bdi}) we see that the dimensions $b_{L,Q}$ do not always increase   
by one when increasing $L$ by one: at every $Nth$ step the dimension stays 
the same:
the dimensions of the $\PQL$ for $L+Q\mod N=0$ and $L+Q\mod N=1$ conincide, 
see e.g. Table 1. 
 
For $\:N=3\;$ the definition (\ref{Bscc}), written explicitly, gives
$$ P_{Q+r}^{(L)}(s)=\frac{t^{-P_a}}{3}\lb (t^2+t+1)^L
   +\om^{Q+r}(t^2+\om^2 t+\om)^L+\om^{-(Q+r)}(t^2+\om t+\om^2)^L\rb. $$

\section{$\Zmb_2$: The Ising case}
In (\ref{zcs}) we have seen three different variables in which to 
write the polynomials: $z,\;c\;$ and $s.\;$ Since for (\ref{EE}) we 
only need the zeros $z_j$ or $c_j$ or $s_j$, the choice which polynomials 
we should prefer can be 
decided by convenience. The range in which the zeros appear 
is dictated by the hermiticity of $\cH^{(s)}$ which requires
\beq -1<c_j<+1 \hs\mbox{or}\hs -\infty<s_j<0 \hx\hs\mbox{or}\hs |z_j|=1.
\end{equation}  
Let us warm up by considering the Ising case $N=2$ for which there are two
charge sectors $Q=0,\:1$ and we have $s=t^2$. From (\ref{Bscc}) Baxter's 
polynomials 
are \beq  P_Q^{(L)}(s=t^2)=\frac{t^{-a_{L,Q}}}{2}\lb(t+1)^L
+(-1)^Q(t-1)^L \rb;\hx a_{L,Q}=L+Q\!\mod 2. \label{szwei}\end{equation}
Their zeros are well known \cite{Ba82}:
$$ s_j=-\mbox{tan}^2\lb\frac{\pi}{L}\lk j-\frac{1-Q}{2}\rk\rb. $$
One easily derives (see e.g. the next section) that for $L$ even
these polynomials satisfy the differential equation (DE) 
$$ 4\,s(s-1)\frac{d^2}{ds^2}\PQL\;-2\,f_Q \:\frac{d}{ds}\PQL\;
             +\,g_Q\:\PQL\:=\:0.  $$
\bea \mbox{with}\hs\hs\; &&f_0=s(2L-3)+1;\hs\hs  g_0=L(L-1);\hs\hs\hx\ny \\
       &&f_1=s(2L-5)+3;\hs\hs g_1=(L-1)(L-2)\hs\hs\hx \label{cz2}\eea
(for $L$ odd interchange $f_0\leftrightarrow f_1$ and $g_0\leftrightarrow g_1$).
These are hypergeometric DEs with the polynomial solutions 
$$\PQL(s)\;\sim\;_2F_1\lk\tfr{Q+1-L}{2}\,,\;\tfr{Q-L}{2}\,;
     \;\:Q+\hal\,;\;s\:\rk\hs\mbox{for}\hx L\;\hx\mbox{even},$$
$$\PQL(s)\;\sim\;_2F_1\lk 1-\tfr{Q+L}{2}\,,\;\hal-\tfr{Q+L}{2}\,;
     \;\:\tfr{3}{2}-Q\,;\;s\:\rk\hs\mbox{for}\hx L\;\hx\mbox{odd}.$$	   
How does the same information appear in terms of the variable $c$ (it is 
the $c_j$ which are directly required in (\ref{EE}))?$\;$ We define 
polynomials $\Pi_Q^{(L)}(c)$ by
\beq \Pi_Q^{(L)}(c)=(c+1)^{b_{L,Q}} P_Q^{(L)}\lk s=\frac{c-1}{c+1}\rk, 
\label{ppi}\end{equation}
where $b_{L,Q}$ was given in (\ref{bdi}). After some algebra, the DE 
(\ref{cz2}) can be rewritten in terms of $\Pi_Q^{(L)}(c)$ and we find
\beq    \frac{d}{dc}\lk(1-c^2)^{Q+\hal}\,\frac{d\Pi_Q^{(L)}}{dc}\rk
        +\frac{(L^2-4Q)(1-c^2)^{Q-\hal}}{4}\;\Pi_Q^{(L)}=0. 
\label{zzwei}\end{equation}
These are Chebyshev-type equations. Their polynomial solutions, correctly 
normalized in accordance with (\ref{ppi}), are:
$$ \Pi_0^{(2k)}(c)=2^k\, T_k (c);\hx\hx \Pi_1^{(2k)}(c)=2^k
U_{k-1}(c);\hx\hx \Pi_{0\atop 1}^{(2k+1)}(c)=2^k(U_k(c)\pm U_{k-1}(c)),$$ 
for $k=0,\,1,\,2,\,\ldots\,\;$ and we use
      $$T_n(x)=\cos{(n\arccos{x})};\hs  
       U_n(x)=\sin{\lb (n+1)\arccos{x}\rb}/\sqrt{1-x^2}.$$ 
As classical orthogonal polynomials the
Chebyshev polynomials satisfy three-term pure recursion relations.  
In terms of the $\Pi_Q^{(L)}$ these become:
\beq \Pi_Q^{(L+4)}-\,4\,c\,\Pi_Q^{(L+2)}+\,4\,\Pi_Q^{(L)}=0. \end{equation}
A main concern later in this talk will be to investigate whether for 
$N\ge 3$ the polynomials $\Pi_Q^{(L)}(c)$ still form orthogonal 
sequences or at least preserve
some features of the Chebyshev polynomials.  

We also consider briefly eqs.(\ref{szwei}) rewritten in terms of the 
variable $z$. The interpolation polynomials $\cF_Q^{(L)}(z)$ are 
related to the   $P_Q^{(L)}(s)$ by 
\beq \cF_Q^{(L)}(z)=
   (z+1)^{2b_{L,Q}}\;P_Q^{(L)}\lk s=\lk\tfr{z-1}{z+1}\rk^2\rk
   \end{equation}
and for $\:N=2\;$ we get  $$L\; \mbox{even}:\hx
2^{1-L}\cF_0^{(L)}=z^L+1;\hx
  2^{1-L}\cF_1^{(L)}=\frac{z^L-1}{z^2-1};\hx\hx L\; \mbox{odd:}\hx
  2^{1-L}\cF_{0\atop 1}^{(L)}= \frac{z^L\mp 1}{z\mp 1}.$$
For $L$ prime these are standard cyclotomic polynomials $Q_k(z)$, 
in some other cases products of $Q_k(z),\,$ e.g.
$\cF_1^{12}(z)=2^{11}Q_3Q_4Q_6Q_{12};\;\cF_0^{13}(z)=2^{12}Q_{13};
\;\cF_1^{13}(z)=2^{12}Q_{26}$.  
The corresponding $\Zmb_3$ interpolating polynomials $\cF_Q^{(L)}(z)$
are more complicated and for them we have not yet discovered 
particular interesting features.

\section{$\Zmb_3$-Polynomials $\;\PQL(s)$}
To derive recursion relations and differential equations it proves convenient
to start with the Baxter polynomials $P_Q^{(L)}(s)$. We shall describe the
details of the derivations for the case $\Zmb_3$. For higher $N$ the
same methods can be used, just the formulae become more involved. Only
occasionally we will quote the result for general $N$.  

\subsection{Recursion relations}
Our starting point is, that equivalent to Baxter's definition (\ref{Bscc}), 
we can define the $\PQL(s)$ via the expansion of the multinomial
$\;(t^{N-1}+t^{N-2}+\ldots+\,t\,+\,1)^L\;$ and collecting the terms with 
powers of $t^a\;$ for each $a\mod N=0,1,\ldots,N-1$.
So, for $N=3$ we write
\beq (t^2+t+1)^L= t^{a_{L,0}}P_0^{(L)}(s)+t^{a_{L,1}}P_1^{(L)}(s)
      +t^{a_{L,2}}P_2^{(L)}(s),  \label{t3}\end{equation}
with $\;a_{L,Q}=2(L+Q)\mod 3\;,$ and the $\PQL$ are required to depend on 
$s=t^3$ only.	  
From (\ref{t3}) we get immediately
\bea \lefteqn{(t^2+t+1)^{L+1}\;=\;
   (t^2+t+1)\lk t^{a_{L,0}}P_0^{(L)}(s)+t^{a_{L,1}}P_1^{(L)}(s)
 +t^{a_{L,2}}P_2^{(L)}(s)\rk}\ny \\ &&=t^{a_{L+1,0}}P_0^{(L+1)}(s)
 +t^{a_{L+1,1}}P_1^{(L+1)}(s)+t^{a_{L+1,2}}P_2^{(L+1)}(s).\hs\hs\hs \eea
Collecting the coefficients of mod 3-powers of $t$, 
and choosing $L$ such that $L\mod 3=0$, this gives 
\beq  \lk\barr{c}P_0^{(L+1)} \\ P_1^{(L+1)} \\ P_2^{(L+1)}\earr\rk=
  \lk\barr{ccc} 1& 1& 1\\ 1& s& 1\\ 1& s& s
 \earr \rk\lk\barr{c}P_0^{(L)}\\P_1^{(L)}\\P_2^{(L)}
 \earr\rk,\hx
\lk\barr{c}P_2^{(L+2)} \\ P_0^{(L+2)} \\ P_1^{(L+2)}\earr\rk=
  \lk\barr{ccc} 1& 1& 1\\ 1& s& 1\\ 1& s& s
 \earr \rk\lk\barr{c}P_2^{(L+1)}\\P_0^{(L+1)}\\P_1^{(L+1)}
 \earr\rk. \label{kr3} \end{equation}
One rotates the column vectors once more
 to get the $P_Q^{(L+3)}$ in terms of the $P_Q^{(L+2)}$.\\
Still taking $L\mod 3=0$, by repeated application of (\ref{kr3})
etc. we obtain 4-term pure recursion relations between polynomials {\it 
of the same} $Q$,$\;$ e.g.:
\bea &&P_0^{(L+3)}\hs\! -\,3\,s\,P_0^{(L+2)}\:+3s(s-1)\;\;P_0^{(L+1)}\:
    +(s-1)^2\,P_0^{(L)}\:=\:0 \ny\\ 
&&P_1^{(L+3)}\hs\hx -3\,P_1^{(L+2)}\;\; +3(s-1)\;\;P_1^{(L+1)}\,
    +(s-1)^2\,P_1^{(L)}\:=\:0 \ny\\
&&P_2^{(L+3)}\hs\! -3\,s\,P_2^{(L+2)}\;\;\; +3(s-1)\;\;P_2^{(L+1)}\,
    +(s-1)^2\,P_2^{(L)}\;=\:0 \ny\\
&&P_Q^{(L+4)}\:-(2s+1)\,P_Q^{(L+3)}\;\;\,+2(s-1)^2\,P_Q^{(L+1)}\:
    +(s-1)^3\,P_Q^{(L)}\:=\:0 \hx\hx \mbox{for}\;Q=1,\:2 \ny\\ 
&&\!\!\!sP_0^{(L+4)}\:-(2s+1)\,P_0^{(L+3)}\:+2s(s-1)^2\,P_0^{(L+1)}\:
    +(s-1)^3\,P_0^{(L)}\:=\:0.	
\label{prec} \eea
In each except the fourth equations pairs of polynomials appear which have 
the same degree. The following recursion is valid for all $L\ge 0$ and $Q$:
\beq P_Q^{(L+9)}-3((s-1)^2+9s)P_Q^{(L+6)}+3(s-1)^4P_Q^{(L+3)}
  -(s-1)^6P_Q^{(L)}=0.\label{prcu} \end{equation}

\subsection{Differential equations}
Differentiating (\ref{t3}), for $L\mod 3=0$ we get:    
\bea \lefteqn{(2\,t+1)L\lk P_0^{(L)}+t^2 P_1^{(L)}+t P_2^{(L)}\rk}\ny\\
 &=&(t^2+t+1)\lb 3\,t^2\lk
 {P_0^{(L)}}'+t^2 {P_1^{(L)}}' +t {P_2^{(L)}}'\rk+2\,tP_1^{(L)}+P_2^{(L)}\rb
 \ny\eea 
 where ${P_Q^{(L)}}'\equiv dP_Q^{(L)}/ds$, and comparing the coefficients 
 of mod 3-powers of $t$:
 \beq 
  3 s(s-1)\lk \barr{c}{\!P_0^{(L)}}'\!\\\!{P_1^{(L)}}'\!
    \\\!{P_2^{(L)}}'\!\earr\rk
 \!=\!\cB\lk \barr{c}\!P_0^{(L)}\!\\\!P_1^{(L)}\!
    \\\!P_2^{(L)}\!\earr\rk;\hx\;\;
 \cB=\lk\barr{ccc}
 2sL& -sL&-sL\\-L& \;2sL\!-2(s-1)& -L\\ -L& -sL&\!\!2sL\!-(s-1)
 \earr \rk\!\!.   \label{dk3} \end{equation} 
Writing (\ref{dk3}) shorthand as $3s(s-1)\cP'=\cB\:\cP$, and differentiating
again we have 
\bea  9s^2(s-1)^2\cP''\!&=&\! \lb\cB^2-3(2s-1)\cB +3s(s-1)\cB'\rb \cP \ny\\ 
     27s^3(s-1)^3\cP'''\!&=&\!\lb \cB^3-9(2s-1)\cB^2+18\,(3s(s-1)+1)\cB
	      +3s(s-1)(2\cB'\,\cB+\cB\,\cB')\right.\ny\\
		  &&\hspace{5cm}\left. -\,18\,s(s-1)(2s-1)\cB'\rb\cP 
\label{derv}\eea
We can use these relations to obtain {\it decoupled}$\:$ DEs for 
the three charge sectors as follows: 
We form the three expressions ($Q=0,1,2$)
\beq 27s^2(s-1)^2{P_Q^{(L)}}'''-27\,s(s-1)\, f_Q\, {P_Q^{(L)}}''
    +3\,g_Q\, {P_Q^{(L)}}'-(L-1)\,h_Q\, P_Q^{(L)}=0. \label{dq3} 
	\end{equation}
which contain 9 coefficients $f_Q,\;g_Q$ and $h_Q$. 
Then using (\ref{derv}) we 
express the ${P_Q^{(L)}}'''$, ${P_Q^{(L)}}''$ and ${P_Q^{(L)}}'$  
in terms of $\PQL$ which gives a matrix equation
${\cal M}\cP=0$. Requiring ${\cal M}$ to be diagonal gives 9 linear 
equations for the coefficients $f_Q,\:g_Q,\:h_Q$. 
Solving these, the result for $L\!\mod 3=0$ is
\bea f_Q&=& s(2L-4-\TQ)+2+\TQ;\hs\hs \TQ=3-Q\mod 3;  \ny\\ 
 g_Q&=&3s(4s-1)L^2-3sL\lb 4(s-1)\TQ+10s-7\rb\ny\\
  &&\hspace*{4cm}+(s-1)\lb 3\TQ^2(s-1)+3\TQ(5s-1)+20s-2)\rb;\ny\\
 h_0&=&\lb (8s+1)L-4(s-1)\rb L;\hs
 h_1\;=\;\lb (8s+1)L-12(s-1)\rb(L-2);\ny\\
 h_2&=&\;(8s+1)L^2-(16s-7)L+6(s-1). \label{dco3}\eea
For the other values of $L\mod 3$ we get the same expressions, 
just for rotated values of $Q\,.\:$ 
E.g. we have the same equation for $L\mod 3=0,\;\;Q=0$ as for 
$L\mod 3=1,\;\;Q=2$.
These DE's (\ref{dq3}) 
are not anti-selfadjoint as one would like them to be. 

For higher $N$ the derivations are completely analogous, just quite lengthy. 
One obtains $Nth$-order DE's for the $\Zmb_N$-polynomials. For $L\!\mod N=0$  
the relation generalizing  (\ref{dk3}) is
$$  Ns(s-1)\frac{d\cP}{ds}= \cB^{(N)}\cP $$                            
\beq \cB^{(N)}=\lk\barr{ccccc}
 \cN & -sL& -sL&-sL&\cdots\\-L&\;\;\cN\!-\!(N\!-\!1)(s\!-\!1)& -L& -L&\cdots\\
-L& -sL&\!\!\cN\!-\!(N\!-\!2)(s\!-\!1)&-L&\cdots\\
 -L& -sL& -sL&\!\!\cN\!-\!(N\!-\!3)(s\!-\!1)\;&\cdots\\ 
 \cdots&\cdots&\cdots&\cdots&\cdots
\earr \rk \label{bmN}\end{equation}
using the abbreviation $\;\cN=(N-1)s\,L\,=\,N\,s\,b_{L,0}\;$ with 
$\;b_{L,Q}\;$ 
being defined in (\ref{bdi}).

\subsection{Approximation of the zeros}
In \cite{Ge99} it was shown that changing the variable $t$ into $\beta$
according to
\beq t=\frac{\sin{(\beta-\tfr{\pi}{N})}}{\sin{(\beta+\tfr{\pi}{N})}},  
\end{equation}
the zeros of the polynomials come approximately equidistant in the interval 
$-\tfr{\pi}{N}<\beta<\tfr{\pi}{N}.$
Without approximations, for $\Zmb_3$ the equation
   $\;\; \cP_{L,Q}(t)=0\;\; $  becomes 
\beq  (2\cos{\beta})^{-L}\;=\;2\cos{\lk L\lb\beta+\frac{\pi}{3}\rb
 +\frac{2\pi Q}{3}\rk}.              \label{trit}\end{equation} 
Neglecting the left-hand side of (\ref{trit})  for $L \gg 1$ (this is 
exponentially good for $|\beta|\ll \pid$ and dangerous only for 
$\;2\cos{\beta}\approx 1,\;$ i.e. $\;\beta\approx \pm\pid\;$) we get 
the approximate solutions 
\beq \beta_k\;\approx\; K-\frac{\pi}{3};\hs K=\frac{6k+2Q-3}{6L}\,\pi;\hs
 k=1,2,\ldots,b_{L,Q}.  \end{equation} or 
\beq c_k\;=\;\frac{1+s_k}{1-s_k}\;\approx -\frac{\SiS^3(K+\pid) 
 -\SiS^3 K}{\SiS^3(K+\pid)+\SiS^3 K}.\end{equation}  
For $N=2$ the analogous approach leads to exact results.

Here we shall not further discuss this useful trigonometric mapping of the 
zeros to the interval $-\tfr{\pi}{N}<\beta<\tfr{\pi}{N}$. Rather we 
focus our attention to the 
rational mapping from $\:-\infty<s<0\;$ to $\:-1<c<1\:$ and study whether 
traces of the orthogonality present in the $N=2$ case survive for $N\ge 3$.

\rnc{\arraystretch}{1.4} 
\begin{table}[t]  \begin{center}
\caption{$\Zmb_3$-Polynomials $\Pic(c)$ for $L\le 8$.} 
{\small\begin{tabular}{llll}\\[-3mm]
\hline $L$ & $\Pi_0^{(L)}$ &$\Pi_1^{(L)}$ &$\Pi_2^{(L)}$ \\ \hline
1& 1& 1& 1\\ 2& $3c+1$ & $3c-1$ & 3\\ 3& $9c^2-5$&$9c+3$&$9c-3$\\
4& $27c^2+18c-5$&$27c^2-11$&$27c^2-18c-5$\\ 5& $81c^3+27c^2-57c-11$&
$81c^3-27c^2-57c+11$&$81c^2-21$\\ 6& $243c^4-270c^2+43$&$243c^3+81c^2-135c-21$
&$243c^3-81c^2-135c+21$\\ 
7&$3^6c^4\!+\!486c^3\!-\!549c^2\!-\!270c\!+\!43\!$&$3^6c^4-702c^2+85$&
 $3^6c^4\!-\!486c^3\!-\!549c^2\!+\!270c\!+\!43\!$\\ 
 8&$3^7c^5+3^6c^4-2754c^3-702c^2$&
$3^7c^5-3^6c^4-2754c^3+702c^2$&$3^7c^4-1782c^2+171$\\[-1mm] 
 &$\hs+711c+85$&$\hs+711c-85$&  \\
\hline \end{tabular}}\end{center}
\end{table}
\rnc{\arraystretch}{1}

\section{The polynomials $\;\Pic(c)$}
The general definition of the $\Pic(c)$ has already been given in (\ref{ppi}). 
For illustration Table 1 lists the $\Pic$ for $L\le 8$. The inversion 
relations (\ref{srcp}) simply imply that
\beq \Pic(c)\:=\:(-1)^{b_{L,Q}}\:\Pi_{N-Q-L}^{(L)}(-c), 
\label{selr}\end{equation}
from which it follows that for $L+2Q\mod N=0$ the polynomials $\Pic$ 
for $N$ odd are functions of $c^2$ only.

\subsection{Recursion relations}

We first derive recursion relations for the case $\Zmb_3$. 
We rewrite the recursive relations obtained for the $\PQL(s)$, (\ref{kr3}) 
in terms of the polynomials $\Pic(c)$,
with the result (taking $L\mod 3=0$):
$$ \lk\barr{c}\Pi_0^{(L+i+1)}\\ \Pi_1^{(L+i+1)}\\ 
 \Pi_2^{(L+i+1)}\earr\rk\;=\;\caC^{(i)}
 \, \lk\barr{c}\Pi_0^{(L+i)} \\ \Pi_1^{(L+i)} \\ \Pi_2^{(L+i)}\earr\rk\!;  
$$
\beq \caC^{(0)}=\lk\barr{ccc} 1\;\;& c_+& c_+ \\1 \;\;&c_- & c_+
   \\ 1\;\;&c_-&c_- \earr\rk\!;
   \hx  \caC^{(1)}=\lk\barr{ccc} c_-& c_+&c_+
   \\ c_-&c_-&c_+\\ 1& 1&1\earr\rk\!; 
\hx\caC^{(2)}=\lk\barr{ccc} c_-& c_+&c_+c_-\\ 1&1&c_+\\ 1& 1&c_-\earr\rk.
\label{mxr}\end{equation} with $\;\;c_\pm=c\pm 1.\;\;$ 
Several interesting relations follow immediately from the particular 
structure of the matrices $\caC^{(i)}:\;$
Still taking $L$ such that $L\mod 3=0$, we have  
\bea \Pi_0^{(L+1)}-\Pi_2^{(L+1)}&=& 2(\Pi_1^{(L)}+\Pi_2^{(L)});\hs\;\:
\Pi_0^{(L+1)}-\Pi_1^{(L+1)}\;=\; 2\Pi_1^{(L)}\ny\\
\Pi_0^{(L+2)}-\Pi_1^{(L+2)}&=& 2\Pi_1^{(L+1)};\hs\hs\hx
\Pi_1^{(L+3)}-\Pi_2^{(L+3)}\;=\; 2\Pi_2^{(L+2)},\label{lr3}  \eea 
which shows that the 9 polynomials of the subset with  
$L\mod 3=0,1,2$ and $Q=0,1,2$ are not independent. There are also
derivative relations between these polynomials, see later eqs.(\ref{sumnu}),
(\ref{derip}).\\[2mm] 
The relations (\ref{mxr}) mix the three charge sectors. In order to
find recursion relations which do not mix the charge sectors $Q$, we form 
the matrices 
$\Csf,\;\Csf^2,\;\Csf^3$:
$$ \Csf\equiv\caC^{(2)}\caC^{(1)}\caC^{(0)}=\lk\barr{ccc} 
   \Pdd_0 &c_+c_-\Pdd_2 &c_+c_-\Pdd_1 \\ \Pdd_1 & \Pdd_0 & c_+\Pdd_2 \\
   \Pdd_2&c_-\Pdd_1& \Pdd_0\earr \rk; \hs \barr{l}\Pdd_0=9c^2-5\\
   \Pdd_1=3c+1\\ \Pdd_2=3c-1.\earr $$
$$ \Csf^2=\lk\barr{ccc} \alpha_2 & c_+c_-\beta_{2+} & c_+c_-\beta_{2-} \\
                \beta_{2-}& \alpha_2 & c_+\beta_{2+} \\
	\beta_{2+}& c_-\beta_{2-}& \alpha_2 \earr\rk; \hs\hs 
	\barr{l} \alpha_2=3^5c^4-270c^2+43 \\			    
             \beta_{2\pm}=3^5c^3\mp 3^4c^2-135c\pm 21 \earr   $$ 
$$ \Csf^3=\lk\barr{ccc} \alpha_3 & c_+c_-\beta_{3+} & c_+c_-\beta_{3-} \\
                \beta_{3-}& \alpha_3 & c_+\beta_{3+} \\
	\beta_{3+}& c_-\beta_{3-}& \alpha_3 \earr\rk; \hx\hx\;\; 
	\barr{l} \alpha_3=3^8c^6-3^7 \cdot 5c^4+3^4\cdot 59c^2-341 \\
\beta_{3\pm}=9(3^6 c^5\mp 3^5c^4 -3^4\cdot 10 c^3\\
			 \hs\hs\pm 3^2\cdot 22 c^2
			 +177c \mp 19) \earr   $$ 
The matrix equation 
$\;\;\Csf^3\:+\,f\,\Csf^2\:+\,g\,\Csf\:+\,h\,\mathbb{I}=0$
contains three independent equations and has the solution 
$f=-27c^2+15=-3\Pi_0^{(3)};\;\;g=48;\;\;h=-64,\;\;$ corresponding 
to the pure recursion relation (valid {\it for all} $\;L\ge 0\;$ {\it and all} 
$\;\,Q=0,1,2\;)$\footnote{If we define $\Pi_{0\atop 1}^{(-1)}=\mp\hal;\;\;
 \Pi_2^{(-1)}=0;\;\; \Pi_{0\atop 2}^{(-2)}=\frac{1}{4};\;\;\Pi_1^{(-2)}
 =-\hal,\;\;$, the relations (\ref{rik}) are valid also for $L=-1,\;-2$.}:  
\beq  
\Pi_Q^{(L+9)}-(27c^2-15)\Pi_Q^{(L+6)}+48\,\Pi_Q^{(L+3)}-64\,\Pi_Q^{(L)}=0.
\label{rik}\end{equation} 
The degrees of the polynomials appearing in this relation increase by two from
the right to the left. If we consider $Q=L\!\mod 3$, then we have only
polynomials in $c^2$, so that the degrees appearing are consecutive in powers
of $z=9c^2$ ("simple sets of polynomials") with integer coefficients. 
These recursion relations
are of simplest type (like those for the Chebyshev polynomials), 
with the coefficients {\it not depending} on $L$. Compare also the 
same equation (\ref{prcu}) for the $\PQL$. 
\\[3mm]
In order to form {\it simple} sets out of the $\Pic(c)$ with the same $Q$ 
having the degree in $c$ increasing in steps of one without repetitions,
every third $L$ must be left out. E.g. for $Q=0$, we can form the sequences 
$L=0,2,3,5,6,8,9,\ldots$ or $L=1,2,4,5,7,8,\ldots$, 
i.e. leaving out either all $L\mod 3=1$ or all $L\mod 3=0$. Analogous 
sequences can be formed for $Q=1$ and for $Q=2$.\\[1mm]
Using again (\ref{mxr}) and solving similar matrix equations, we find the
following 4-term pure recursion relations: 
For all $L\ge 0$ with $L\mod 3=0$ we get 
\bea   && \Pi_Q^{(L+4+j)}\:+(1-3c)\,\Pi_Q^{(L+3+j)}\:+8\,\Pi_Q^{(L+1+j)}
 \:+8\,\Pi_Q^{(L+j)}=0;\ny \\  &&
 \hspace*{4cm}\mbox{for}\hx j=0,1,2\hx\mbox{with}\hx Q=4-j,\;2-j \mod 3; 
 \label{mnrc} \\[2mm]
    && \Pi_Q^{(L+5+j)}\:+(1-9c)\,\Pi_Q^{(L+3+j)}\:+(8-12c)\,\Pi_Q^{(L+2+j)}
 \:+8\,\Pi_Q^{(L+j)}=0;\ny \\  &&
 \hspace*{4cm}\mbox{for}\hx j=0,1,2\hx\mbox{with}\hx Q=3-j,\;2-j \mod 3.
 \hs\hs \label{nnrc}\eea
Classical orthogonal polynomials must have 3-term recursion relations 
(see e.g. \cite{Chih}). The
separation of the polynomials $\PQL$ for a fixed $Q$ into two sets: the
ones selected in (\ref{mnrc},\ref{nnrc}) and the ones left out, reminds 
of Konhauser biorthogonal polynomials \cite{Konh} (KBOP). 
KBOPs have 4-term pure recursion relations if the first set consists of the 
polynomials in $c$ and second set of polynomials in $c^2$. 
In some cases of (\ref{mnrc}) this is what we just did, (the 
$\Pi_Q^{(L+Q)}$ are functions of $c^2$), but not in all cases: for $j=2$
in (\ref{mnrc}) $\Pi_{2,0}^{L+1}$ are left out and these depend on $c$. 

We can apply the same technique to find recursion relations also for
higher $\Zmb_N$ polynomials. Generally, we find $N+1$-term pure recursion relations.
For $\Zmb_4$ the result analogous to (\ref{rik}) is
\beq \Pi_Q^{(L+16)}-4\Pi_0^{(4)}\Pi_Q^{(L+12)}-128
 (14c^2-17)\Pi_Q^{(L+8)}-2048c\Pi_Q^{(L+4)}+4096\Pi_Q^{(L)}=0. 
\label{rkvier}\end{equation}
with $\Pi_0^{(4)}=64c^3-56c.\;\;$  For $\Zmb_5$:
\beq \Pi_Q^{(L+25)}-5\,\Pi_0^{(5)}\,\Pi_Q^{(L+20)}+a_2\Pi_Q^{(L+15)}
  +a_3\Pi_Q^{(L+10)}+a_4\Pi_Q^{(L+5)}+a_5\Pi_Q^{(L)}=0. \end{equation}
\bea \mbox{with}\hx\hx&&-5\Pi_0^{(5)}=-3125\,c^4\,+3750\,c^2\,-705;\hs a_2=
  -2^6\,5\,(375\, c^2-383);\hs\;\ny\\ 
  && \hx\hx  a_3=-2^{10}(5^4c^2-45\cdot 13);\hs 
	 a_4=2^{16}\cdot 5;\hs a_5=-2^{20}.\ny \eea  
There are also $N+1$-term recursion relations analogous to 
(\ref{mnrc}),$\:$(\ref{nnrc}) for sequences formed of $\Pic(c)$ with
consecutive degrees in $c$. Generally, for this, certain every $Nth$ $L$
must be left out. If these were KBOPs, for having $N+1$-term pure 
recursions, there should be subsets of polynomials depending on $c^N$.
However, this is not the case: for any $N$ there are only subsets 
depending on $c^2$ because of (\ref{selr}).  

One might think that orthogonal polynomials which usefully 
approximate the $\Pic$ could be obtained recursively, neglecting the 
last term in (\ref{rik}).
Using monic polynomials $\widetilde{\Pi}_Q^{L}$, this looks promising 
because eq.(\ref{rik}) becomes
$$\widetilde{\Pi}_Q^{(L+9)}=\lk c^2-\frac{5}{9}\rk\widetilde{\Pi}_Q^{(L+6)}
 -\frac{16}{243}\widetilde{\Pi}_Q^{(L+3)}
 +\frac{64}{19683}\widetilde{\Pi}_Q^{(L)}.   $$
with the last coefficient looking small. However, new polynomials 
$\widetilde{R}_0^{(L)}$ defined through (considering $L\mod 3=0,\;Q=0$):
$$\widetilde{R}_0^{(L+6)}=\lk c^2-\frac{5}{9}\rk\widetilde{R}_0^{(L+3)}
 -\frac{16}{243}\widetilde{R}_0^{(L)}$$
with $\;\widetilde{R}_0^{(0)}=3;\;\;\widetilde{R}_0^{(3)}=c^2-\frac{5}{9}$ 
don't seem to be useful, since the $\widetilde{R}_0^{(L)}$ have zeros which 
for increasing $L$ spread out of the interval $-1<c<+1$, violating the 
hermiticity of the hamiltonian. The method using eq.(\ref{trit}) does not
suffer from this deficiency.

\subsection{Differential equations for the $\Pic$(c)}
The derivative relation (\ref{bmN}) can be rewritten for the $\Pic$, giving
e.g. for $L\mod N=0$:
\beq \lk\barr{c}
 {\Pi_0}'\\c_+{\Pi_1}'\\ c_+{\Pi_2}'\\c_+{\Pi_3}'\\
 \cdots\earr\rk=
\frac{1}{Nc_+c_-}\lk \barr{ccccc}0&Lc_-&Lc_-&Lc_-&\cdots\\
   Lc_+&-Nc_+\!+\!2\!&Lc_+&Lc_+&\cdots\\
   Lc_+&Lc_-&-Nc_+\!+\!4\!&Lc_+&\cdots\\
   Lc_+&Lc_-&Lc_-&-Nc_+\!+\!6\!&\cdots\\
   \cdots&\cdots&\cdots&\cdots&\cdots        \earr\rk\!
   \lk\barr{c}\Pi_0\\c_+\Pi_1\\c_+\Pi_2\\c_+\Pi_3\\
   \cdots           \earr\rk \label{derr}\end{equation}
where the primes denote differentiation $d/dc$ and we omitted the 
superscripts $(L)$ on the $\Pi_Q$.
The first component of this equation gives the nice 
simple relation
\beq {\Pi_0}'=\frac{L}{N}\sum_{Q=1}^{N-1}\Pi_Q. \label{sumnu}\end{equation}
We now concentrate on the case $N=3$.
From relations analogous to (\ref{derr}) we get the following simple 
derivative relations:
\bea 3c_-{\Pi_0}'&=&L(\Pi_1+\Pi_2)-2\Pi_0;\hs
     3c_+{\Pi_2}'\;=\;L(\Pi_0+\Pi_1)-2\Pi_2 \ny\\
   && 3(c^2-1){\Pi_1}'\;=\;L(c_+\Pi_2+c_-\Pi_0)-2c\Pi_1.\label{derip} \eea 
Using these and similar relations, after considerable algebra, we get the DE
\bea 27(c^2\!-\!1)\lb (c^2\!-\!1)\frac{d^3\Pi_Q^{(L)}}{d\,c^3}
    \;+\;\alpha(c)\frac{d^2\Pi_Q^{(L)}}{d\,c^2}\rb\!\!\! &-& 
	3\lb 3(c^2\!-\!1)L(L+1)+\beta(c)\rb\frac{d\Pi_Q^{(L)}}{d\,c}\ny\\[-2mm]
  &+&\gamma\,(L,c)\:\Pi_Q^{(L)}=0.\;\label{Dd3}  \eea
\rnc{\arraystretch}{1.7}
\begin{table}[ht]
\begin{center}
\caption{Coefficients appearing in the DE (\ref{Dd3}) for the 
$\Zmb_3$-polynomials $\Pic(c)$}
\begin{tabular}{ccccl} \hline
$L\mod 3$& $Q$ & $\hx\alpha(c)\hx$&$\beta(c)$&$\hs\hs\hs\gamma(L,c)$\\ 
  \hline
0&0& $4c$ & $-18c^2+10$ & $-2cL^2(L+3)$\\  
0&$1\atop 2$& 
   $7c\pm 1$ & $-90c^2\mp 24c+34$ & $ -2cL^2(L+3)-3(3c\pm 1)(L(L+1)-6)$\\ 
1&1& $6c$ & $-60c^2+28 $ & $-2c\lb L^2(L+6)+ 3L-10\rb $ \\
1&$2\atop 0$&
   $6c\mp 2$&$-60c^2\pm 36c+16$&$-2c\lb L^2(L+6)+3L-10\rb\pm 6(L(L+1)-\!2)$\\
2&2& $8c$& $ -126c^2+46 $&$-2c\lb L^2(L+9)+6L-56\rb$ \\
2&$0\atop 1$
    &$5c\pm 1$&$-36c^2\mp 12c+16$ &$-c\lb L^2(2L+9)+3L-4\rb\mp 3L(L+1)$\\    
\hline  \end{tabular}\end{center}
\end{table}
\rnc{\arraystretch}{1}
We collect the coefficients in Table 2. 
Observe that these DE for the 
$\Pic(c)$ are much simpler than the corresponding eqs.(\ref{dq3}) for the
$\PQL(s)$: there the coefficients $f_Q$ of the second derivatives 
${\PQL}''$ (\ref{dco3}) were $L$-dependent,
 which here is not the case. However, although in (\ref{Dd3}) the third and
 second derivative terms can be combined into a total derivative, we don't 
 see how to put the whole DE into an anti-selfconjugate form.

\subsection{Separation property of the zeros.}
We want to collect further evidence that the polynomials $\Pic$ violate
features which are crucial for classical orthogonal polynomials.
Table 3 shows the zeros of the polynomials $\Pi_0^{(L)}$ for low $L$. 
Fully satisfied is the confinement of the zeros $c_j$ to the interior 
of the basic interval $-1<c_j<+1$ (this must be so because of the
hermiticity of the hamiltonian). However, the separation property of
the zeros of polynomials sequences successive in $c$ can not be satisfied: 
If we take out
the $L=3,6,9,\ldots$-polynomials as we did in (\ref{mnrc}) for $j=1$
with $Q=4-j$ then the zeros $.78444;\;.77622$ for $L=5,7$ 
are in wrong order. Alternatively, taking out $L=4,7,11,\ldots$ as in
(\ref{mnrc}) for $j=2$ with $Q=2-j$ then $.95835;\;.93330$ for $L=6,8$
are in wrong order too. Considering the polynomials in $c^2$ 
only ($Q=0$ and $L=3,6,9,\ldots$ as in (\ref{rik})) gives the correct 
separation of the $c_j^2$.

\rnc{\arraystretch}{1.5}
\begin{table}[ht]
\caption{Zeros $c_j$ of the $Z_3$-polynoms $\Pi_0^{(L)}(c)$
showing the separation property violation at the upper corner near $c=1$.}
{\small\begin{center}\begin{tabular}{r rrr rrr rrr}\hline 
 $L$& \mco{9}{c}{$\hs\hs c_j\hq (\:j\:=\:1,\,\ldots,\,[\,2L/3\,]\:)$ }\\ \hline
 2& &&  &&&\mco{2}{c}{ -.33333 }&&\\
 3& &&  &&& -.74536& .74536 &&\\
 4& &&&  &\mco{3}{c}{ -.87766$\;\;\:\;$ .21100} &&\\
 5& &&&& -.93203 & -.18575 & .78444 && \\  
 6& &&& \mco{5}{c}{-.95835$\;\;\;$-.43894$\hx\;\;\;$.43894$\hx\;\;$.95835} &\\ 
  \hline
 7& &&&   -.97264& -.60038& .13013& .77622 &&\\
 8& && \mco{6}{c}{   -.98106$\;\;\;$-.70632 $\;\;\;\;$-.11267 $\hx
              $.53341 $\hx\;$.93330} &\\
 9& && -.98634&-.77818&-.29702& .29702&.77818&.98634&\\
 10 &&\mco{7}{c}{$\!\!\!$-.98983$\hx\;$-.82848$\hx\;$-.43636$\hx\;$ .09047$\hx
        \;       $ .58910$\;\;\;\;$ .90843}&\\
 11 &$\hq\;$& -.99222&-.86467&-.54241&-.08238& .39948& .77767& .97120&\\  
 12 & \mco{9}{c}{$\;\;\;$-.99392$\hq\!\!$-.89136$\,\hx$-.62401$\hx\;
              $-.22459$\hx\;$ .22459$\hx\;$
           .62401$\;\;\;\;$ .89136$\;\;\:\hx$ .99392}\\ \hline
\end{tabular}\end{center}}
\end{table}

\subsection{The $\Pic$ expressed in terms of Jacobi polynomials:\\
Partial orthogonality}

In order to find out whether the $\Pic(c)$ satisfy some generalized 
kind of orthogonality, we looked for the eventual weight functions. 
The definition interval is certainly $-1<c<+1$ and as a simple guess
one may try Jacobi weights
$$\:w_{\alp,\beta}(c)=(1-c)^\alp \:(1+c)^\beta.\:$$ 
Since for $\Zmb_2$ the $\Pic$ are Chebyshev polynomials, i.e. there
$\alp=\beta=\pm 1/N$, we tried whether e.g. the $\Zmb_3$ polynomials
are related to $\alp,\; \beta=\pm \frac{1}{3}\; \pm\frac{2}{3}$-Jacobi 
polynomials. Indeed, expanding the $\Pic(c)$ in terms of
Jacobi polynomials $P_n^{(\alp,\beta)}(c)$ and playing around with 
several choices of $a$ and $b$ in $\alp=a/N$ and $\beta=b/N$ we find 
remarkable features which we report now. 
\begin{table}[ht]
\rnc{\arraystretch}{1.7}
\caption{Expansion vectors (\ref{jacv}) of the 
$\Zmb_3$-polynomials $6^{-[2L/3]}\:\Pic$ in terms of Jacobi polynomials. 
Note the many vanishing components $a_0,\:a_1,$ etc.}
\begin{center}\begin{tabular}{clll} \hline
\mco{4}{c}{$6^{-[2L/3]}\:\Pic(c)\hx$ for $\hx\Zmb_3\hx$}\\ \hline
$L$  &  $\hs\hx Q=0$  &  $\hs Q=1$  &  $\hs Q=2$  \\ \hline
2& $\Big[ 0,\frac{1}{2}\:\Big]_{\frac{1}{3},-\frac{1}{3}}$ & 
  $\Big[ 0,\frac{1}{2}\:\Big]_{-\frac{1}{3},\frac{1}{3}}$ & 
  $\Big[\frac{1}{2}\:\Big]_{\frac{1}{3},\frac{2}{3}}$ \\
3& $\Big[ 0,\frac{1}{9},\frac{1}{3}\:\Big]_{-\frac{2}{3},-\frac{1}{3}}$ & 
   $\Lb 0,\frac{1}{4}\RBb$& $\Lb 0,\frac{1}{4}\RBa$  \\  
4& $\Lb 0,0,\frac{1}{2}\RCb$&$\Lb 0,\frac{1}{4},\frac{1}{2}\RBa$&
   $\Lb 0,0,\frac{1}{2}\RCa $\\
5& $\Lb 0,-\frac{1}{18},0,\frac{3}{20}\RBb$ &  
   $\Lb 0,-\frac{1}{18},0,\frac{3}{20}\RBa$ & 
   $\Lb 0,\frac{1}{30},\frac{3}{20}\RDa$\\          
6& $\Lb 0,0,-\frac{1}{45},\frac{1}{35},\frac{3}{35}\RAa$ &  
   $\Lb 0,0,0,\frac{3}{40}\RBb$ & $\Lb 0,0,0,\frac{3}{40}\RBa$\\  
7& $\Lb 0,0,-\frac{1}{36},0,\frac{9}{70}\RCb$&
   $\Lb 0,0,-\frac{1}{63},\frac{3}{40},\frac{9}{70}\RBa$&
   $\Lb 0,0,-\frac{1}{36},0,\frac{9}{70}\RCa$\\
8& $\Lb 0,0,0,-\frac{1}{45},0,\frac{1}{28}\RBb$&
   $\Lb 0,0,0,-\frac{1}{45},0,\frac{1}{28}\RBa$&
   $\Lb 0,0,-\frac{1}{168},\frac{1}{105},\frac{1}{28}\RDa$\\
9& $\Lb 0,0,0,\frac{-1}{756},\frac{-1}{105},\frac{1}{154},
    \frac{3}{154}\RAa\!\!\!$& 
   $\Lb 0,0,0,\frac{-1}{360},0,\frac{1}{56}\RBb$&
   $\Lb 0,0,0,\frac{-1}{360},0,\frac{1}{56}\RBa$\\  
10&$\Lb 0,0,0,0,-\frac{1}{77},0,\frac{9}{308}\RCb$&
   $\Lb 0,0,0,\frac{-1}{360},\frac{-1}{110},\frac{1}{56},\frac{9}{308}\RBa
    \!\!\!$&
   $\Lb 0,0,0,0,\frac{-1}{77},0,\frac{9}{308}\RCa$\\
\hline\end{tabular}\end{center}
\end{table}
\rnc{\arraystretch}{1.5}
\begin{table}[ht]
\caption{Expansion (\ref{chev}) of $\Zmb_4$-polynomials 
$2^{-[3L/4]}\:\Pic$ in terms of Chebyshev polynomials. The same components
appear (partly shifted) e.g. for $7_{0+1},\:7_{2+3}$ and $8_{1+3}$.}
\begin{center}\begin{tabular}{clcl} \hline
$L_{Q_1+Q_2}$&$\hx 2^{-[3L/4]}\:\hal\,
                      (\Pi_{Q_1}^{(L)}(c)+\Pi_{Q_2}^{(L)}(c))$  & 
$L_Q$&$\hx 2^{-[3L/4]}\:\Pic(c)$  \\ \hline
$1_{1+2}$ & $\Lb 1\Rb_U$   & $2_1$ & $\Lb 0,2 \Rb_T$    \\
$2_{0+2}$ & $\Lb 0,1 \Rb_U$       & $2_3$ & $\Lb 2 \Rb_U$   \\ \hline
$3_{0+1}$ & $\Lb 0,0,2 \Rb_T$     & $4_0$ & $\Lb 0,-1,0,2\Rb_T$ \\
$3_{2+3}$ & $\Lb 0,2 \Rb_U$       & $4_2$  & $\Lb -1,0,2 \Rb_U$ \\
$4_{1+3}$ & $\Lb 0,0,2 \Rb_U$     & $5_{0+3}\!\!\!$ & $\Lb 0,1,0,4 \Rb_U$ 
 \\ \hline
$5_{1+2}$ & $\Lb 0,-3,0,4 \Rb_U$  & $6_1$ &  $\Lb 0,0,-2,0,8 \Rb_T$ \\
$6_{0+2}$ & $\Lb 0,0,-3,0,4\Rb_U$ & $6_3$ & $\Lb 0,-2,0,8 \Rb_U$ \\ 
 \hline
$7_{0+1}$ & $\Lb 0,0,0,-4,0,8\Rb_T$ & $8_0$ & $\Lb 0,0,1,0,-8,0,8 \Rb_T$\\
$7_{2+3}$ & $\Lb 0,0,-4,0,8\Rb_U$   & $8_2$ & $\Lb 0,1,0,-8,0,8 \Rb_U$ \\
$8_{1+3}$ & $\Lb 0,0,0,-4,0,8\Rb_U$     & 
$9_{0+3}\!\!\!$&$\Lb 0,0,-3,0,-4,0,16\Rb_U$ \\ \hline
$9_{1+2}$ & $\Lb 0,0,5,0,-20,0,16\Rb_U$  & $10_1$ & 
$\Lb 0,0,0,2,0,-24,0,32 \Rb_T$ \\
$10_{0+2}$& $\Lb 0,0,0,5,0,-20,0,16\Rb_U$& $10_3$ & 
$\Lb 0,0,2,0,-24,0,32 \Rb_U$ \\ 
\hline
$11_{0+1}$ & $\Lb 0,0,0,0,6,0,-32,0,32\Rb_T \hs\hq$    & 
$12_0$ & $\Lb 0,0,0,-1,0,18,0,-48,0,32\Rb_T$ \\
$11_{2+3}$ & $\Lb 0,0,0,6,0,-32,0,32\Rb_U$      & 
$12_2$ & $\Lb 0,0,-1,0,18,0,-48,0,32\Rb_U$ \\
$12_{1+3}$ & $\Lb 0,0,0,0,6,0,-32,0,32\Rb_U$    & 
$13_{0+3}\!\!\!$&$\Lb 0,0,0,5,0,-8,0,-48,0,64\Rb_U$ \\ \hline
\end{tabular}\end{center}
\end{table}

\begin{table}[ht]
\caption{Expansion vectors for some $\Zmb_5$- and $\Zmb_6$-polynomials. 
Since in several cases we don't show all components but just the first 
and last ones, in the column
$dim[\:]$ we give the lengths of the component vectors.}
\begin{center}\begin{tabular}{ccl} \hline
$L_Q$&$dim[\:]$&$\hs 5^{-[L/5]}\:\Pi_Q^{(L)}(c)\hx$for $\hx\Zmb_5\hx$\\ \hline 
$4_2$& 3 &$\Lb 0,0,\frac{250}{3}\REa$ \\
$5_1$& 4 &$\Lb 0,-8,0,50 \RGa $\\[1mm]
$6_1$& 5 &$\Lb 0,0,\frac{-500}{7},0,\frac{1000}{7}\REa$ \\[2mm]
$6_{0\atop3}$& 5 &$\Lb 0,\mp 16,0,\pm 150,\frac{1000}{7}\RFa$ \\[2mm]
$9_2$&7&$\Lb 0,0,\frac{4400}{21},0,\frac{-260000}{77},0,
                          \frac{1250000}{231}\REa$\\
$14_2$& 11&$\Lb 0,0,0,0,\frac{-1792000}{143},0,\frac{8 \cdot 10^7}{561},0,
  \frac{-92\cdot 10^8}{24453},0,\frac{125\cdot 10^8}{46189}\REa$ \\
$19_2$& 15&$\Lb 0,0,0,0,\frac{-2048000}{221},0,\frac{128 
 \cdot 10^5}{21},0,\ldots,
 0,\frac{-55 \cdot 10^{12}}{2028117},0,\frac{25 \cdot 10^{11}}{200583}\REa$ \\
$24_2$& 19&$\Lb 0,0,0,0,0,0,\frac{2301952000}{2737},0,\ldots,0,
\frac{-688 \cdot 10^{15}}{420756273},0,\frac{5 \cdot
                          10^{16}}{90751353}\REa$\\   
\\[-2mm] \hline
$3_3$& 2 & $\Lb 0,25 \RFa$ \\[1mm] 
$8_3$& 6 & $\Lb 0,0,0,\frac{-2000}{3},0,\frac{125000}{63}\RFa$\\
$13_3$&10& $\Lb 0,0,0,\frac{-9600}{11},0,\frac{74 \cdot 10^5}{273},0,
    \frac{-8 \cdot 10^8}{7293},0,\frac{25 \cdot 10^7}{2431}\RFa$\\  
$18_3$& 14&$\Lb 0,0,0,0,0,\frac{1056 \cdot 10^5}{1729},0,
  \ldots,0,\frac{25 \cdot 10^{10}}{52003}\RFa$ \\
$38_3$& 30&$\Lb 0,0,0,0,0,0,0,0,0,\frac{478150656 \cdot 10^5}{268801},0,
  \ldots,0,\frac{625 \cdot 10^{25}}{375840831244263}\RFa$  \\
\hline $L_Q$&$dim[\:]$&$\hs 6^{-[L/6]}\:\Pi_Q^{(L)}(c)
   \hx$for$\hx\Zmb_6\hx$\\ \hline 
$3_4$ & 2 & $\Lb 0,36\Rb_{\frac{1}{6},-\frac{1}{6}}$\\[1mm]
$4_3$ & 3 & $\Lb 0,0,144\Rb_{\frac{1}{3},-\frac{1}{3}}$\\[1mm]
$10_3$& 8 & $\Lb 0,0,0,8736,0,\frac{-5045760}{91},0,\frac{8957952}{143}
               \Rb_{\frac{1}{3},-\frac{1}{3}}$\\[1mm]
$6_0$ & 6 & $\Lb 0,\frac{43}{3},0,-90,0,81\Rb_T$ \\[1mm]
$6_3$ & 5 & $\Lb \frac{43}{3},0,-90,0,81\Rb_U$ \\
\hline
\end{tabular}\end{center}\end{table} 

We use the standard definitions given e.g. in Szeg\"o \cite{Szeg} 
with
$$ P_n^{(\alp,\,\beta)}(1)=\lk n+\alp\atop \alp \rk;\hs 
   P_0^{(\alp,\,\beta)}(c)=\:1\,;
   \hs P_n^{(\alp,\,\beta)}(c)\:=\:(-1)^n\:P_n^{(\beta,\,\alp)}(-c).$$ 
We present our results in Table 4 using the notation 
\beq  \Lb\:a_0,\:a_1,\:\ldots,a_n\:\Rb_{\alp,\beta}\;\equiv 
   \sum_{k=0}^n\;a_k\,P_k^{(\alp,\,\beta)}(c).\label{jacv}\end{equation}
($n$ can be seen from the number of arguments). Surprisingly, when
making a convenient chioce of $\alp$ and $\beta$, with
increasing $L$ an increasing number of the lowest components of the vectors
$[a_0,a_1,\ldots]_{\alp,\beta}$ turns out to be exactly zero. So,
indeed our $\Zmb_3$-polynomials are related to Jacobi $P_k^{(\alp,\beta)}$.
Defining a scalar product 
\beq (\Pic,\:\Pi_{Q'}^{(L')})_{\alp,\beta}\:=\:\int_{-1}^1\,
 dc\:(1-c)^\alp(1+c)^\beta\:\Pi_Q^{(L)}(c)\:\Pi_{Q'}^{(L')}(c) 
\end{equation} 
by inspection of Table 4  for $\:N=3,\:$
we conjecture the "partial orthogonality" 
\beq (\Pi_0^{(3k)},\:\Pi_0^{(3k')})_{-\frac{2}{3},-\frac{1}{3}}\:=\:0
 \hq \mbox{for} \hq 2k+1\le k'. \label{orr}\end{equation}
because the expansion vector of $\Pi_0^{3k}$ has $2k+1$ components and
for $k\le 35$ we have checked that its $k$ lowest components are zero.    
Interestingly, the first non-zero components are very small and increase
strongly before reaching a plateau level around the $kth$ component:
E.g. for $\Pi_0^{105}$ we have $a_0=\ldots=a_{34}=0;\;a_{35}=-0.0007395;\;
a_{38}=14.6665;\;a_{52}=-3.186\cdot 10^9$. It is a simple consequence of
the reflection relation (\ref{selr}) that for fixed $L$ always one pair 
of $\Zmb_3$ charge sectors differs only by the interchange of 
$\alp$ and $\beta$. \\[1mm]
Several other partial orthogonality relations can be inferred
from other vanishing first components in Table 4.     

For the $\Zmb_4$-polynomials we find that these show a similar partial 
orthogonality if expanded in terms of
Chebyshev polynomials $U_n(c)$ and $T_n(c)$, i.e. for $\alp=\beta=\pm \hal.$ 
Here using the normalisation of the Chebyshev polynomials leads to
much simpler coefficients than using $P_n^{(\pm\hal,\pm\hal)}(c)$, so we 
define  \beq  \Lb\:a_0,\:a_1,\:\ldots,a_n\:\Rb_U\;\equiv 
         \sum_{k=0}^n\;a_k\,U_k(c), \label{chev} \end{equation}
analogously for $T.\;$  Table 5 lists results for $L\le 13$.   
Observe the remarkable symmetry between pairs of sectors obtained by 
replacing $T_k$ by $U_{k-1}$ (e.g. for $\;\Pi_{1+2}^{(L+1)}\;$ and 
$\;\Pi_{0+2}^{(L+2)}\;$ taking $L\!\mod 4=0$) or, 
replacing $T_k$ by $U_{k}$ for $\;\Pi_{0+1}^{(L)}\;$ and 
$\;\Pi_{1+3}^{(L+1)}\;$ (taking still $L\!\mod 4=0$). 
These relations are not at all simple if expressed in terms of 
the $\Pic(c)$. 

We conclude these observations giving in Table 6 few results for $N=5$ 
and $N=6$: For $N=5$ we find partial orthogonality due to vanishing
first components with $\alp=-\beta=\pm \frac{2}{5}$ and 
$\alp=-\beta=\pm \frac{1}{5}$ according to the sectors $L\!\mod 5$ and $Q$ 
considered. For $N=6$ both Chebyshev- and e.g. 
$\alp=-\beta=\pm \frac{1}{6}$-Jacobi weights are useful. As for $N=4$, various
symmetries show up, e.g. for $N=6,\;L\!\mod 6=0$ exchanging $Q=0$ with 
$Q=3$ corresponds to exchanging the Chebyshev polynomials 
$T_k$ and $U_{k-1}$. 
A convenient method to perform expansions in terms of Jacobi 
polynomials is given in \cite{Carl}, Sec.7.2.

\section{Conclusions}
We have studied various properties of the polynomials, which via 
their zeros determine the energy eigenvalues of hamiltonians satisfying
Onsager's algebra. Three different versions have been considered with
the zeros in the hermitian case on the negative real axis, the interval 
$(-1,+1)$, and the unit circle, respectively. The most simple
properties emerged in the second case, the polynomials $\Pic(c)$
which for $\Zmb_2$ are Chebyshev polynomials. 
For the general $\Zmb_N$ we argue that the polynomials considered 
have no shorter than
$N+1$-term pure recursion relations. This excludes that the $\Pic$
are classical orthogonal polynomials. Konhauser biorthogonality (which 
allows 4-term and higher recursion relations) is not seen to be a 
property of the $\Pic$. However, 
the expansion in terms of Jacobi polynomials reveals a very remarkable
partial orthogonality with respect to Jacobi weight functions.
The deeper meaning of the latter remains to be studied.
There are more general definitions of biorthogonal polynomials in the 
literature \cite{Br92}, e.g. of Iserles and N{\o}rsett \cite{Is87}
and Van Iseghem \cite{vI87}. For hypergeometric functions $_2F_2$ there
are four-term recursion relations, see e.g. \cite{Rain} Chap.14. However, 
these have a much more complicated structure than the recursions 
which we found for the $\Pic$. 
More work is needed to clarify the possible relevance of these latter
structures. 

\subsection*{Acknowledgements}
GvG is grateful to Michael Baake, Harry Braden and Nikita Slavnov
for fruitful discussions. He thanks the Institute of Mathematics, 
Academia Sinica, Taipei for kind hospitality, INTAS-97-1312 and 
the National Center for Theoretical Sciences of the Tsing Hua 
University in Hsinchu, Taiwan, for support. 

{\small
}

\end{document}